\documentclass[conference]{IEEEtran}
\IEEEoverridecommandlockouts
\usepackage{cite}
\usepackage{amsmath,amssymb,amsfonts}
\usepackage{algorithmic}
\usepackage{graphicx}
\usepackage{textcomp}
\usepackage{xcolor}
\usepackage{hyperref}
\def\BibTeX{{\rm B\kern-.05em{\sc i\kern-.025em b}\kern-.08em
    T\kern-.1667em\lower.7ex\hbox{E}\kern-.125emX}}

\begin{document}
%
\title{Towards end-to-end F0 voice conversion based on Dual-GAN with convolutional wavelet kernels}

\author{\IEEEauthorblockN{Clément Le Moine}
\IEEEauthorblockA{IRCAM, CNRS, Sorbonne Université \\
STMS Lab, \\
Paris, France \\
lemoineveillon@ircam.fr}
\and
\IEEEauthorblockN{Nicolas Obin}
\IEEEauthorblockA{IRCAM, CNRS, Sorbonne Université \\
STMS Lab, \\
Paris, France \\
nicolas.obin@ircam.fr}
\and
\IEEEauthorblockN{Axel Roebel}
\IEEEauthorblockA{IRCAM, CNRS, Sorbonne Université \\
STMS Lab, \\
Paris, France \\
axel.roabel@ircam.fr}}


%


\maketitle

\begin{abstract}
This paper presents a end-to-end framework for the F0 transformation in the context of expressive voice conversion. A single neural network is proposed, in which a first module is used to learn F0 representation over different temporal scales and a second adversarial module is used to learn the transformation from one emotion to another. The first module is composed of a convolution layer with wavelet kernels so that the various temporal scales of F0 variations can be efficiently encoded. The single decomposition/transformation network allows to learn in a end-to-end manner the F0 decomposition that are optimal with respect to the transformation, directly from the raw F0 signal.
\end{abstract}


%
\IEEEpeerreviewmaketitle

\section{Introduction}
\label{sec:intro}

Fundamental frequency (F0) is an essential acoustic feature in human speech communication and human-human interactions. As a key feature of speech prosody, F0 plays an important role in every aspect of speech communication: it conveys linguistic information (F0 helps to clarify the syntactic structure of an utterance or is used for semantic emphasis), para-linguistic information such as emotion or social attitude, and is even a part of the speaker identity through his speaking style.
Accordingly, generative F0 modeling can be extremely useful in the fields of text-to-speech \cite{wang2018style}, voice identity conversion (VC) \cite{Kaneko_2018_CGANVC}, and  expressive voice conversion \cite{Luo_2017}, by allowing a direct and parametric control of F0 to manipulate the expressivity of a voice (such as speaking style or emotions).
By nature, F0 variations occur over different temporal scales each associated with specific functions, ranging from micro-variations to macro-contours such as accentuation, emotions, and modalities. To cover these specificities of F0 modelling, stylization methods \cite{Teu08, Obi18} and multi-level modelling \cite{Obi11c, Yin16, Wang2017-xu} have been proposed.

Notably in VC, generative models such as Gaussian mixture models \cite{VeauxGMM} and LSTM-based Sequence-to-Sequence models \cite{robinson2019} were used to learn F0 transformations from neutral to expressive speech. Lately, various works focused on the use of Continuous Wavelet Transform (CWT), as an intermediary representation of the F0, on which Generative Adversarial Networks (GAN) models such as Dual-GAN \cite{Luo_2019}, Cycle-GAN \cite{zhou2020_CG}, VAW-GAN \cite{zhou2020_VAWG} or VA-GAN \cite{luo_chen_takiguchi_ariki_2019} are trained to learn transformations. A majority of those models is learnt on parallel data and by emotion pairs, which allows to learn a direct mapping between two different emotional versions of an utterance while preserving a fixed and controlled linguistic content. 

A promising approach called CWT Adaptive Scales (CWT-AS) was proposed by Luo et al. \cite{Luo_2019}. The CWT computes a decomposition of the F0 signal over wavelet kernels which allows a representation of F0 over different temporal scales \cite{Ming15}, with various application in expressive voice conversion \cite{Ming15, Ming2016-tp, Luo_2017, Luo_2019, zhou2020_VAWG}. F0 modelling with CWT has been specified more recently upgraded with the possibility to compute the decomposition on arbitrary linguistic scales (e.g., phoneme, syllable, word, and utterance as described in \cite{Luo_2017}). An Adaptive-Scale (AS) algorithm \cite{Luo_2019} is described to select an optimal CWT representation for each pair of emotions, by selecting the scales that maximize in average the distance between the emotions in the CWT space.  

From these selected scales, the CWT decomposition of the F0 contours is computed. Finally, the transformation function between each pair of emotion is learned from those representation using a Dual-GAN. 
Though this approach appears promising, it suffers from two main limitations: 1) The scale selection is only based on the maximization of the distance between the emotions, but ignores their reconstruction ability of the F0 signal. This may lead to poor F0 reconstruction which in turn would degrade the quality and the naturalness of the transformation; 2) The CWT-AS decomposition of the F0 signal and the dual-GAN are optimized independently which constitutes a bottleneck for training. Consequently, the CWT decomposition may not be optimal in the sense of the dual-GAN objective. 

To overcome those limitations, we propose a end-to-end architecture to learn efficiently F0 transformation between emotions. The proposed neural architecture brings together the F0 decomposition and the dual-GAN into a single network, so that the CWT decomposition is optimized in the  sense of the dual-GAN objective, and combining separation and reconstruction losses of the resulting decomposition. An application to the voice conversion of social attitudes shows that the proposed approach significantly improves the quality of the transformation by comparison with the CWT-AS approach.

\section{PROPOSED METHOD}
\label{sec:meth}

In this section, we introduce our proposal based on CWT-AS \cite{Luo_2019} and show how it differs by integrating the F0 stylization part on top of the transformation learning process which we refer to as end-to-end method for voice f0 conversion in \ref{Framework}. Our contributions' concepts and technical details are given in \ref{WKCE} and \ref{Model}.


\subsection{Framework overview}
\label{Framework}

\begin{figure*}[h]
    \centering
    \includegraphics[width=\textwidth]{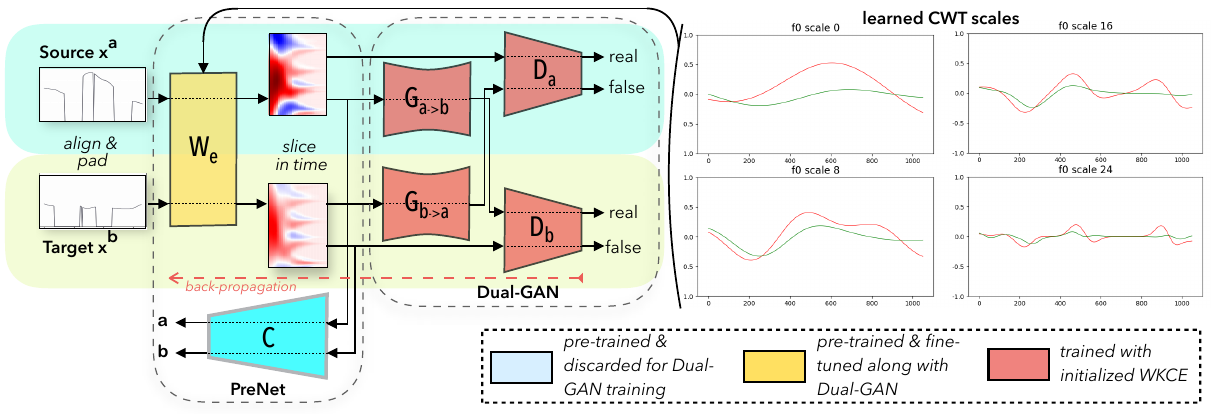}
    \caption{End-to-end neural architecture for F0 voice conversion. On the right, F0 decomposition over 4 of the learned scales obtained for the source (red) and target (green) expressivities.}
    \label{dual_gan}
\end{figure*}

As our proposed VC system requires parallel data, sets of utterances $X^{a}$ and $X^{b}$ respectively relative to expressivity $a$ and $b$ are considered. A pair of utterances is then sampled and F0 sequences, source $\mathbf{x}^{a}$ and target $\mathbf{x}^{b}$, are extracted. Aside from the expressivity, each utterance in a pair has the same content (linguistic content, speaker identity). The source and target F0 are given to what we called a Wavelet Kernel Convolutional Encoder (WKCE) denoted $W_{e}$. A classifier, denoted $C$, whose objective is to predict the expressivity is fed with WKCE outputs. As shown in Figure \ref{dual_gan}, these two modules must be seen as a pre-network ($pN$) for Dual-GAN ($DG$) that can be pre-trained as well as trained along with Dual-GAN forming an end-to-end system for f0 conversion. 

\subsection{Wavelet Kernel Convolutional Autoencoder}
\label{WKCE}

As a multiscale modelling method, CWT is entirely fitting when trying to represent both long and short-term dependencies, prosody is influenced by. As CWT can only be applied to continuous functions, a simple linear interpolation between voiced F0 segments is needed to obtain a continuous phrase-related F0 function which can then be sampled in a vector $\mathbf{x} \in [0, 1]^{T}$. 

Our WKCE performs convolutions between the F0 signal $\mathbf{x}$ and a wavelet kernel based on a mother wavelet $\boldsymbol{\psi_{s}} \in \mathbb{R}^{T}$ defined for a time vector $\mathbf{t} \in \mathbb{R}^{T}$ as

\begin{equation}
\small
\boldsymbol{\psi}_{s} = \frac{2\pi^{-\frac{1}{4}}}{\sqrt{3}}(1-\left(\frac{\mathbf{t}}{s}\right)^{2})e^{-\frac{1}{2} \left(\frac{\mathbf{t}}{s}\right)^{2}}
\end{equation}

Considering a wavelet kernel depending on $N$ learnable parameters $s$ that control the width of each wavelet composing the kernel, the contribution  $\mathbf{h}_{\mathbf{x}}^{s}$ of the temporal level $s$ to the F0 signal $\mathbf{x}$ is the convolution between $\mathbf{x}$ and $\boldsymbol{\psi}_{s}$. Therefore, feed with $\mathbf{x}$, our WKCE module will output $W_{e}(\mathbf{x}) = [\mathbf{h}_{\mathbf{x}}^{s_{0}}, ..., \mathbf{h}_{\mathbf{x}}^{s_{N}}] \in \mathbb{R}^{N \times T}$. If we denote $W_{r}$ the reconstruction operation, then the reconstructed signal $\hat{\mathbf{x}}$ is given by

\begin{equation}
\small
	\hat{\mathbf{x}} = W_{r}(W_{e}(\mathbf{x})) = \frac{d_{j}\sqrt{d_{t}}}{C_{d} Y_{0}}\sum_{i=0}^{N-1} \mathbf{h}_{\mathbf{x}}^{s_{i}} + \mathbf{\bar{x}}
	\label{cwt_rec}
\end{equation}

with $\mathbf{\bar{x}}$ the average of $\mathbf{x}$, $d_{t}=1.2$, $d_{j}=0.125$, $C_{d}=3.541$ and $Y_{0} = 0.867$ (for details, see \cite{Torrence1998APG}).

If we denote $\mathbb{E}$, the mathematical expectation and consider $\mathbf{x}^{a}$ and $\mathbf{x}^{b}$ sampled from source and target distributions $P(\mathbf{x}^{a})$ and $P(\mathbf{x}^{b})$ respectively, this module can be trained for reconstruction objective with respect to $L_{1}$ loss formulated as follows

\vspace*{-\baselineskip}
\vspace{0.8mm}

\begin{equation}
    \small
    \begin{split}
    L_{rec} = &\mathbb{E}_{\mathbf{x}^{a} \sim P(\mathbf{x}^{a})}(||W_{r}(W_{e}(\mathbf{x}^{a}))) - \mathbf{x}^{a}||_{1}) +\\                     
            &\mathbb{E}_{\mathbf{x}^{b} \sim P(\mathbf{x}^{b})}(||W_{r}(W_{e}(\mathbf{x}^{b})) - \mathbf{x}^{b}||_{1}) 
    \end{split}
    \label{rec}
\end{equation}

A constraint of classification on the CWTs latent space can be added, $W_{e}$ and $C$ are trained with respect to $L_{cl}$, the cross-entropy (CE) loss between the predicted source expressivity $\hat{a} = C(W_{e}(\mathbf{x}^{a}))$ and the true value $a$ summed with the CE between $\hat{b}$ and $b$.


\begin{equation}
\small
\begin{split}
    L_{cl} = &\mathbb{E}_{\mathbf{x}^{a} \sim P(\mathbf{x}^{a})}[a*C(W_{e}(\mathbf{x}^{a}))] +                       \mathbb{E}_{\mathbf{x}^{b} \sim P(\mathbf{x}^{b})}[b*C(W_{e}(\mathbf{x}^{b}))] \\
              + &\mathbb{E}_{\mathbf{x}^{b} \sim P(\mathbf{x}^{b})}(1-a)[1 - log(C(W_{e}(\mathbf{x}^{a})))] \\
              + &\mathbb{E}_{\mathbf{x}^{a} \sim P(\mathbf{x}^{a})}(1-b)[1 - log(C(W_{e}(\mathbf{x}^{b}))))] 
\end{split}
\label{adv}
\end{equation}


\subsection{Model}
\label{Model}

In this paper we focus on a specific GAN network called Dual-GAN which is capable of learning a mapping between parallel pairs of data. This network is based on two concepts: 1) Adversarial learning \cite{NIPS2014_5423}, which is to train a generative model to find a solution in a min-max game between two neural networks, called as generator $G$ and discriminator $D$. 2) Dual supervised learning \cite{Xia_2017} which is to train the models of two dual tasks simultaneously exploiting the probabilistic correlation between them to regularize the training process. Combining those breakthroughs allows to take advantage of the GAN's ability to produce realistic transformations as well as the significant improvements due to dual supervised learning.

This second point implies that both forward and inverse transformations, respectively $G_{a \rightarrow b} : (W_{e}(\mathbf{x}^{a}), z^{a}) \rightarrow \mathbf{x}^{b}$ and $G_{b \rightarrow a} : (W_{e}(\mathbf{x}^{b}, z^{b}) \rightarrow \mathbf{x}^{a}$, are learned jointly, where $z^{a}$ and $z^{b}$ are random independant noises provided in the form of dropout at each layer of $G_{a}$ and $G_{b}$. A first loss $L_{a \leftrightarrow b}$ is required to train $G_{a \rightarrow b}$, $G_{b \rightarrow a}$ and $W_{e}$.


\begin{equation}
\small
    \begin{split}
    L_{a \leftrightarrow b} = &\mathbb{E}_{(\mathbf{x}^{a}, \mathbf{x}^{b}) \sim P(\mathbf{x}^{a}, \mathbf{x}^{b})}(||W_{r}(G_{a \rightarrow b}(W_{e}(\mathbf{x}^{a}))) - \mathbf{x}^{b}||_{1}) \\                     
            + &\mathbb{E}_{(\mathbf{x}^{a}, \mathbf{x}^{b}) \sim P(\mathbf{x}^{a}, \mathbf{x}^{b})}(||W_{r}(G_{b \rightarrow a}(W_{e}(\mathbf{x}^{b}))) - \mathbf{x}^{a}||_{1}) 
    \end{split}
    \label{rec}
\end{equation}

In the same time, $D_{a}$ discriminates between converted outputs $\hat{X^{b}}$ of $G_{a \rightarrow b}$ and real samples of domain $X^{b}$, $D_{b}$ does analogously the same to complete the adversarial mechanism. The adversarial loss $L_{ADV}$ is required to train $G_{a \rightarrow b}$, $G_{b \rightarrow a}$, $D_{a}$, $D_{b}$ and $W_{e}$


\begin{equation}
\small
\begin{split}
    L_{adv} = &\mathbb{E}_{\mathbf{x}^{a} \sim P(\mathbf{x}^{a})}[D_{a}(W_{e}(\mathbf{x}^{a}))] +                      \mathbb{E}_{\mathbf{x}^{b} \sim P(\mathbf{x}^{b})}[D_{b}(W_{e}(\mathbf{x}^{b}))] \\
              + &\mathbb{E}_{\mathbf{x}^{b} \sim P(\mathbf{x}^{b})}[1 - log(D_{a}(G_{b \rightarrow a}(W_{e}(\mathbf{x}^{b}))))] \\
              + &\mathbb{E}_{\mathbf{x}^{a} \sim P(\mathbf{x}^{a})}[1 - log(D_{b}(G_{a \rightarrow b}(W_{e}(\mathbf{x}^{a}))))] 
\end{split}
\label{adv}
\end{equation}

A third constraint called Dual loss is added so as to strengthen the intrinsic connection between $G_{a \rightarrow b}$ and $G_{b \rightarrow a}$, it can be understood as a regularization of the process.

\begin{equation}
\small
\begin{split}
    L_{dual} = \mathbb{E}_{(\mathbf{x}^{a}, \mathbf{x}^{b}) \sim P(\mathbf{x}^{a}, \mathbf{x}^{b})}(||&W_{e}(\mathbf{x}^{a})*G_{a \rightarrow b}(W_{e}(\mathbf{x}^{a})) \\
    - &W_{e}(\mathbf{x}^{b})*G_{b \rightarrow a}(W_{e}(\mathbf{x}^{b}))||_{1})    
\end{split}
\label{adv}
\end{equation}

Therefore two final losses can be formulated for pre-Net pretraining and proper Dual-GAN training, respectively $L_{pN}$ and $L_{DG}$ with $\alpha$, $\beta$, $\lambda$ and $\gamma$ respectively weighting reconstruction, classification, transformation and dual objectives.

\vspace*{-\baselineskip}
\vspace{0.8mm}

{\footnotesize
\begin{align}
\small
L_{pN} &= \alpha L_{rec} + \beta L_{cl} \\
 L_{DG} &= \lambda L_{a \leftrightarrow b} +  L_{adv} + \gamma L_{dual}
\end{align}
}%

\section{Experiments}
\label{sec:exp}


\subsection{Dataset}

For our experiments we used the freely available speech database Att-HACK \cite{LeMoine_sp}. The database comprises 25 speakers interpreting 100 utterances in 4 social attitudes : friendly, distant, dominant and seductive, later denoted FR, DIST, DOM and SED respectively. With 3 to 5 repetitions each per attitude for a total of around 30 hours of speech, the database offers a wide variety of prosodic strategies in the expression of attitudes. 
Prosodic features conveying expressivity has been shown to be speaker dependent \cite{Sisman_18}, for this reason, two speakers were selected and used independently for both training and validation : a female (F08) and a male (M07), representing almost 400 utterances each. The \textit{train/valid} split was 80/20 \% and has been done linguistically. 


\subsection{Implementation details}

\subsubsection{Input pipeline}

We extracted fundamental frequency from the speech signal by using SWIPEP algorithm \cite{Cam07}. All F0 sequences are sampled to 1ms (as recommended in \cite{Luo_2019}), passed to log(F0) and a linear interpolation has been processed between voiced segments. For each pair, a mapping between syllables starting and ending times has been done to align source and target. Once pairs are aligned syllable-wise, F0 sequences are padded with zeros up to a value $T=4000$.


\subsubsection{Architecture design}

Our WKCE denoted $W_{e}$, with $N=32$ learnable scales, has been implemented as a custom layer. A constraint of growth has been added on the range of scales to ensure the continuity of the learned CWTs. The output of $W_{e}$ of shape [32, 4000] is unpadded and temporally sliced to form batches of shape [batch\_size, 32, 512]. 


The classifier is built using convolutional blocks composed of three convolutional layers starting from 32 up to 128 filters. Each block uses 3 × 3 convolutions with \textit{ReLU} activation, a dropout of 0.2, padding mode same and pooling operations using both strides and max pooling 2D with values 2 and 4, respectively, reducing features and time. Those blocks are followed with a flatten layer and two dense layers with respectively, 1000 units and a \textit{ReLU} activation, and 2 units and a \textit{softmax} activation.

Two configurations of our pre-network $pN$ can be distinguished : \begin{itemize}
    \item  \textit{config\_A} : $pN = \{W_{e}\}$ learns the CWT scales regarding the CWT reconstruction objective ($\alpha=1$, $\beta=0$)
    \item \textit{config\_B} : $pN = \{W_{e} + C\}$ learns the CWT scales regarding both the CWT reconstruction and the CWT related attitude classification objectives ($\alpha=10$, $\beta=1$)
\end{itemize}

The architecture of the Dual-GAN itself as well as the contribution of each module, $G_{a \rightarrow b}$, $G_{b \rightarrow a}$, $D_{a}$ and $D_{b}$, in training process are taken from \cite{Luo_2019}.

\vspace{0.8mm}

\subsubsection{Training procedure}

$W_{e}$ is being fed with batches of size 1 so as to allow the Dual-GAN to process on phrase related batches. $pN$ is pre-trained minimizing $L_{pN}$ depending on the considered configuration. CWTs are learned with respect to voiced segments by using a voicing binary mask of 1 (voiced) and 0 (unvoiced). Then $\{DG + W_{e}\}$ is trained minimizing $L_{DG}$ with $\lambda=5$ and $\gamma=15$. For \textit{config\_B}, the classification scores after pre-training were passed as sample weights for Dual-GAN training. ADAM optimizer with $0.0001$ as learning rate has been used. All codes are written in Python-Tensorflow 2.1, the baseline has been re-implemented.

\section{RESULTS AND DISCUSSIONS}
\label{sec:res}

\subsection{Objective Evaluation}
\label{sssec:obj}

\begin{figure}[h!]
    \centering
    \includegraphics[width=0.49\textwidth]{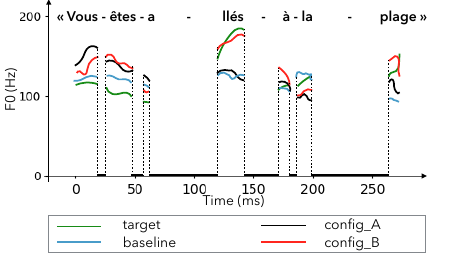}
    \caption{Example of F0 transformation from distant to dominant for speaker M07 for \textit{baseline} and ours \textit{config\_A} and \textit{config\_B}.}
    \label{conv}
  \end{figure}

Considering 4 attitudes, we evaluated 12 transformations (6 forward, 6 invert), the overall results are shown in Table \ref{RMSE}, using the RMSE between 1) the original F0 and the reconstruction obtained from its representation, 2) the converted F0 and the corresponding target F0. An example of conversion is depicted in Figure \ref{conv}.

First of all, we observe that the proposed end-to-end system for f0 voice conversion outperforms the traditional \textit{baseline}, in average over all categories, for both configurations. Secondly our most sophisticated configuration \textit{config\_B} achieves slightly better results than \textit{config\_A}.

\begin{table}[h]
\centering
\begin{tabular}{c|cc}
    & \multicolumn{2}{c}{\textbf{RMSE (Hz)}} \\ 
    \hline
    \textbf{Models} & \textbf{Reconstruction} & \textbf{Transformation} \\ 
    \hline
    \textit{baseline} & 17.32 & 21.71 \\
    \textit{config\_A} & 9.16 & 19.15 \\
    \textit{config\_B} & 13.4 & \textbf{18.83} \\
\end{tabular}
\caption{A comparison of the RMSE results of the \textit{baseline}, \textit{config\_A} and \textit{config\_B} for reconstruction and transformation}
\label{RMSE}
\end{table}

\vspace{-5.5mm}

\vspace*{-\baselineskip}

\subsection{Subjective Evaluation}
\label{sssec:sub}

\begin{figure}[h!]
    \centering
    \includegraphics[width=0.35\textwidth]{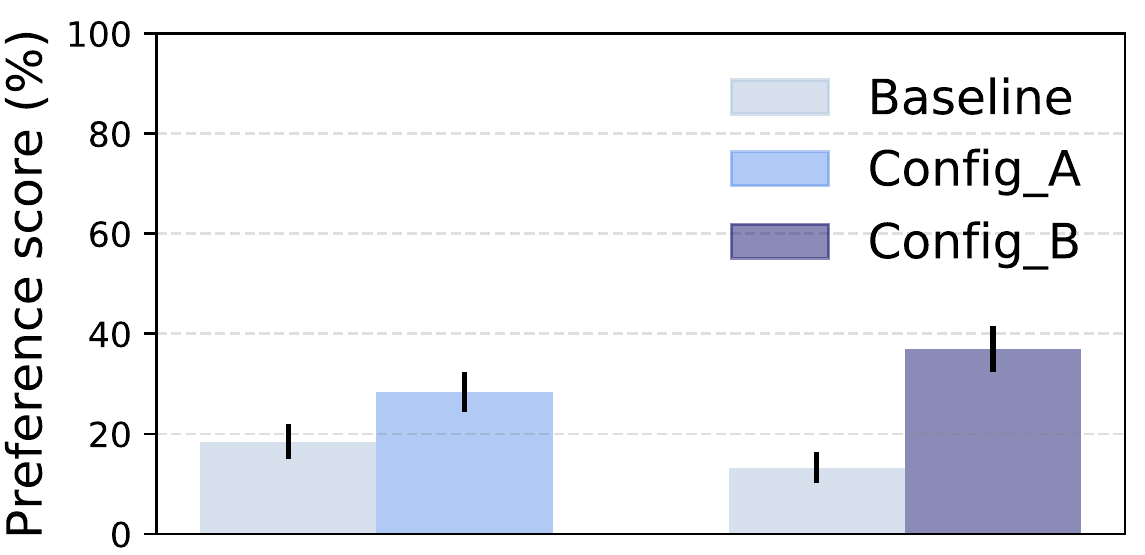}
    \caption{The XAB preference results with 95\% confidence interval between the \textit{baseline} and ours \textit{config\_A} and \textit{config\_B} regarding attitude similarity.}
    \label{XAB}
  \end{figure}

We further conducted a listening experiment to compare the two proposed configurations with the baseline CWT-AS in terms of attitude similarity. We perform XAB test to assess the emotion similarity by asking listeners to choose between two converted utterances (\textit{baseline} and one of our configurations) the one which sounds more similar to the original target. The overall results are reported in Figure \ref{XAB}. Our proposed system in \textit{config\_B} outperforms the baseline for each of the considered transformations. The results for \textit{config\_A}, slightly lower, still outperforms the \textit{baseline}. 

\vspace*{-\baselineskip}

\vspace*{-\baselineskip}

\subsection{Scales distributions}

\begin{figure}[h]
    \centering
    \includegraphics[width=0.5\textwidth]{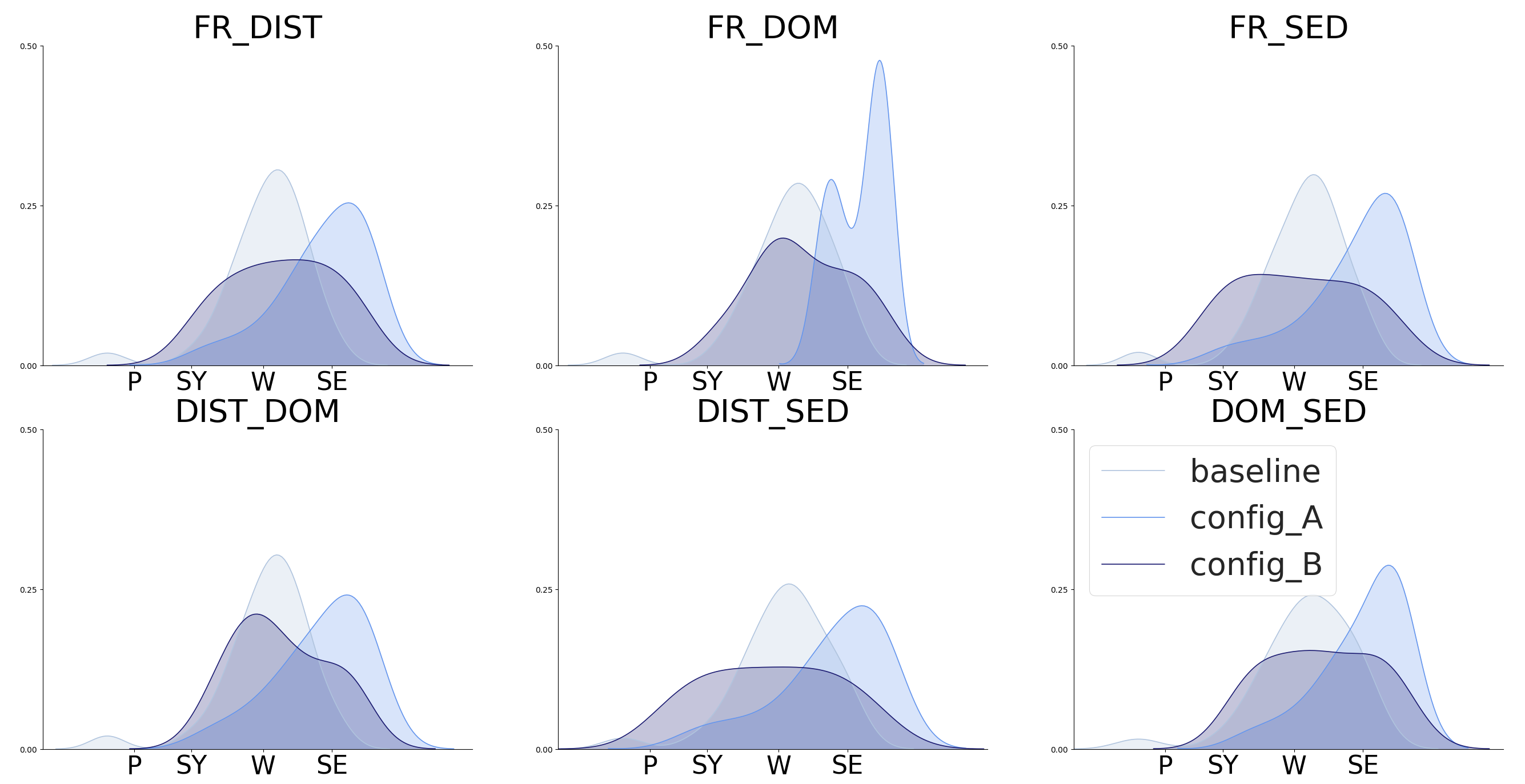}
    \caption{CWT scale distributions for the three considered algorithms: \textit{baseline}, \textit{config\_A} and \textit{config\_B}. For interpretation, P, SY, W and SE markers respectivelly denote average duration of the phone, syllable, word and sentence linguistic units.}
    \label{dist}
\end{figure}

This part provides an a posteriori comparison of the F0 scales distribution as obtained by the description presented in parts \ref{sssec:obj} and \ref{sssec:sub}. Each transformation (forward and backward) between a pair of attitudes is associated with a set of temporal scales that are used to compute the CWT representations used for the conversion. Consequently, each transformation can be described by a distribution of the temporal scales that are used to convert the F0 optimally.  Figure \ref{dist} presents the distribution selected by the baseline CWT-AS algorithm and learned by our proposed contribution, as obtained for speaker F08 for the six pairs of attitudes.

The best performance being obtained with the \textit{config\_B} of the proposed contribution, we further investigated by comparison of the F0 scales distribution. 
First, one can clearly observe that the temporal distribution of the \textit{config\_B} is wider than the others, the transformation covering a wide range of temporal scales from the micro variations over the phonemes to the global contours of the sentence. Additionally, the distributions associated with the \textit{baseline} and the \textit{config\_A} appear mostly independent with respect to the transformation pair, while the distribution associated with the \textit{config\_B} tend to be more varied depending on the transformation pair. This suggests that the\textit{config\_B} may adapt more efficiently to the specificities of each pair



\vspace*{-\baselineskip}
\vspace{1.5mm}

\vspace{-4.5mm}
\section{Conclusion}
\label{sec:con}


\vspace{-1.5mm}

In this paper, we propose a end-to-end framework for F0 transformation in the context of expressive voice conversion, bringing together the F0 decomposition in different temporal levels and its transformation in a single network. Both objective and subjective evaluations showed our method can achieve better performance than the baseline. We aim at generalizing for multi-speaker F0 conversion and to avoid pair-learning by building an expressive embedding. An online page featuring conversion examples is available at \href{http://recherche.ircam.fr/anasyn/VC_demo/index.html}{http://recherche.ircam.fr/anasyn/VC\_demo/index.html}.

\section*{Acknowledgements}
\label{sec:akn}
This research is supported by the MoVe project: ‘‘MOdelling of speech attitudes and application to an expressive conversationnal agent'', and funded by the Paris Region Ph2D grant.



%

\bibliographystyle{IEEEbib}
\bibliography{main}

\end{document}